\documentclass[floatfix,amssymb,prl,twocolumn,superscriptaddress,preprintnumbers,nofootinbib,aps]{revtex4-2}

\usepackage{amssymb,amsmath,mathtools,graphicx,microtype,bm,booktabs}
\usepackage[dvipsnames]{xcolor}
\definecolor{linkcolor}{rgb}{0.0,0.3,0.5}
\usepackage[unicode,colorlinks=true,linkcolor=linkcolor,citecolor=linkcolor,urlcolor=linkcolor,pdfusetitle]{hyperref}
\usepackage[T1]{fontenc}
\usepackage[utf8]{inputenc}

\newcommand{\ustc}{Department of Astronomy, University of Science and Technology of China, Hefei, Anhui 230026, China}
\newcommand{\ustcs}{School of Astronomy and Space Sciences, University of Science and Technology of China, Hefei, Anhui 230026, China}
\newcommand{\ucas}{School of Astronomy and Space Science, University of Chinese Academy of Sciences, Yuquan Road, Beijing 100049, China}
\newcommand{\ihep}{Key Laboratory for Particle Astrophysics, Institute of High Energy Physics, Chinese Academy of Sciences, Yuquan Road, Beijing 100049, China}

\begin{document}

\title{Rapid Dark Growth of Seed Black Holes in Self-Interacting Dark Matter}

\author{Yu Rong}
\email{corresponding author: rongyua@ustc.edu.cn}
\affiliation{\ustc}
\affiliation{\ustcs}
\author{Shuang-Nan Zhang}
\affiliation{\ihep}
\affiliation{\ucas}
\author{Jian-Min Wang}
\affiliation{\ihep}
\affiliation{\ucas}
\author{Junxian Wang}
\affiliation{\ustc}
\affiliation{\ustcs}

\date{September 17, 2026}

\begin{abstract}
The rapid emergence of massive black holes in the early Universe, possibly related to Little Red Dots, challenges conventional growth from light seeds.  Dark-matter accretion offers a distinct route, and self-interacting dark matter (SIDM) makes it possible by behaving as a collisional fluid.  We show that gas inflow can drive strong local contraction of an SIDM halo, assembling a dense reservoir around a pre-existing black-hole seed.  The seed then enters a rapid dark-growth phase and can gain several orders of magnitude in mass within less than $1$ Myr before transitioning to slower accretion.  We self-consistently follow this process by solving the spherical Euler equations with SIDM self-gravity, collisional heat transport, and an absorbing black-hole sink, rather than imposing an accretion history.  The growth depends most strongly on halo mass, with weaker sensitivity to inflowing gas fraction, nuclear size, and seed mass.  Gas-inflow-driven SIDM accretion therefore provides an efficient pathway for producing massive black holes in the early Universe.
\end{abstract}

\maketitle

\noindent\textbf{\emph{Introduction.---}}
The rapid appearance of massive black holes in the early Universe is sharpened by JWST discoveries of compact red sources, or Little Red Dots, at redshifts $z\gtrsim4$--8.  Many show broad lines and spectral energy distributions consistent with dense, obscured, rapidly growing nuclei \cite{Greene2024LRD,Matthee2024LRD,Kocevski2024LRDLF,Akins2025LRD,Labbe2024UltraluminousLRD,Hviding2025RUBIESLRD,InayoshiMaiolino2024DenseGas}.  Together with $z\gtrsim6$ quasars, these systems sharpen a basic timing problem: light black-hole seeds growing only through radiatively efficient gas accretion have little time to reach the observed masses.  Massive seeds, direct-collapse episodes, and near- or super-Eddington gas accretion are leading proposed solutions \cite{MadauRees2001PopIII,BrommLarson2004FirstStars,Volonteri2010SMBH,Mayer2014DirectCollapse,Mayer2023CosmoDCBH,Jeon2025DCBHLRD,PacucciFerraraKocevski2026DCBH}.  Each, however, depends on uncertain gas delivery through angular momentum and feedback.  It is therefore important to test physically distinct channels for growing the seed itself.

One such possibility is dark-matter accretion.  Previous idealized work showed that if dark matter behaves as a collisional fluid around a black hole and an isothermal reservoir supplies it indefinitely, the resulting Bondi accretion rate can be extremely large, allowing the black hole to undergo rapid growth \cite{Bondi1952,Hu2006SIDMacc,BinneyTremaine2008}.  Let $c_s$ be the fixed one-dimensional velocity scale of that reservoir.  If the dark-matter supply is sufficiently collisional, the ideal Bondi supply rate is
\begin{equation}
\dot M_{\rm ideal}=\frac{2c_s^3}{G},
\label{eq:infinite_supply}
\end{equation}
where $G$ is Newton's constant and an overdot denotes a time derivative.  This rate is very large, so a seed supplied by such a reservoir can undergo rapid growth.  The Bondi radius is $r_B=GM_{\rm BH}/c_s^2$ \cite{Bondi1952}.  The dark-matter mean free path at this scale is $\lambda=[\rho(r_B)(\sigma/m)]^{-1}$, where $\rho(r_B)$ is the dark-matter density and $\sigma/m$ is its scattering cross section per unit mass.
When $\lambda<r_B$, collisions keep the dark matter coupled across the Bondi region and allow the high-efficiency fluid-accretion mode to persist.  When $\lambda>r_B$, particles can cross the Bondi region with fewer collisions and are captured less efficiently, so the accretion rate decreases.  We characterize the transition with
\begin{equation}
{\rm Kn}\equiv\frac{\lambda}{r_B}=1,
\label{eq:knudsen}
\end{equation}
where ${\rm Kn}=1$ marks the change from rapid fluid accretion to slower accretion.  For the ideal singular-isothermal reservoir, this condition gives the corresponding saturation mass $M_{\rm sat}$ and ideal-growth timescale $\tau_{\rm sat}$:
\begin{equation}
M_{\rm sat}=\frac{(\sigma/m)c_s^4}{2\pi G^2},\qquad
\tau_{\rm sat}=\frac{M_{\rm sat}-M_0}{\dot M_{\rm ideal}}
\simeq\frac{(\sigma/m)c_s}{4\pi G}.
\label{eq:saturation}
\end{equation}
Thus, a seed with initial mass $M_0\ll M_{\rm sat}$ reaches the saturation scale on the timescale $\tau_{\rm sat}$ and then enters a slower dark-matter growth regime; its mass need not stop increasing.

The numerical scale is striking.  For $\sigma/m=3\,{\rm cm^2\,g^{-1}}$ and $c_s=25\,{\rm km\,s^{-1}}$, characteristic of a $10^9M_\odot$ halo at $z=8$, a $100M_\odot$ seed reaches $10^6M_\odot$ in $0.14$ Myr and the formal saturation mass $M_{\rm sat}=2.09\times10^6M_\odot$ in $\tau_{\rm sat}=0.28$ Myr.  An ideal collisional dark reservoir could therefore bypass the usual gas-accretion bottleneck.

The first difficulty is microscopic.  In ordinary collisionless cold dark matter, the effective scattering cross section is far too small for this fluid-like supply to operate; particles do not efficiently dissipate energy or replenish the central loss cone \cite{BinneyTremaine2008,Hu2006SIDMacc}.  Self-interacting dark matter (SIDM), for which the nonzero $\sigma/m$ permits collisional transport, makes the ideal mechanism physically conceivable \cite{SpergelSteinhardt2000SIDM,BalbergShapiroInagaki2002,TulinYu2018SIDM}.

SIDM alone is nevertheless not enough.  The second difficulty is macroscopic: even a cuspy Navarro--Frenk--White (NFW) halo is finite, and its central density cannot maintain the singular, isothermal, indefinitely replenished reservoir assumed in Eq.~(\ref{eq:infinite_supply}).  As dark matter flows inward, the central supply is depleted and readjusted.  Thus the problem is to identify an astrophysical process that concentrates SIDM into a dense, finite reservoir and to calculate the resulting supply rather than adopt the ideal rate \cite{DasKalita2024SIDM,Sabarish2025SIDMspikes}.

Merger-driven gas inflow can provide that process.  High-redshift mergers and compaction events can carry $10^8$--$10^9M_\odot$ of gas into parsec-scale nuclei \cite{Mayer2014DirectCollapse,Mayer2023CosmoDCBH,Zwick2022MergerDCBH}.  Their deepening gravitational potential compresses the SIDM; sufficiently compact baryonic components can raise the central SIDM density strongly and can trigger local gravothermal collapse \cite{Elbert2016SIDMDisks,Despali2019SIDMBaryons,Sameie2018SIDMBaryons,FengYuZhong2021Seed}.  Gas inflow can therefore build the central SIDM reservoir that an unperturbed NFW halo lacks.

This is the problem we solve here.  Gas inflow compresses the SIDM, collisional transport redistributes it, and black-hole accretion drains and rearranges the same reservoir, so all three processes must be evolved together.  We calculate how a low-mass seed black hole grows by accreting SIDM during gas inflow, determine which nuclear configurations produce a rapid-growth episode, and compare the resulting finite-halo histories with the ideal infinite-supply limit.

\noindent\textbf{\emph{Time-dependent model.---}}
We solve the spherical conservation laws for SIDM mass, radial momentum, and energy, including gravity and collisional heat transport.  All fluid variables below refer to SIDM: $\rho$ is its mass density, $u$ its bulk radial velocity, $P=\rho v^2$ its effective pressure from the one-dimensional random velocity dispersion $v$, and $E=P/(\gamma-1)+\rho u^2/2$ its internal-plus-bulk-kinetic energy density, with adiabatic index $\gamma=5/3$ \cite{LyndenBellEggleton1980,BalbergShapiroInagaki2002,KodaShapiro2011}:
\begin{align}
\partial_t\rho+\frac{1}{r^2}\partial_r(r^2\rho u)&=0,\nonumber\\
\partial_t(\rho u)+\frac{1}{r^2}\partial_r[r^2(\rho u^2+P)]
 &=\frac{2P}{r}-\rho g,\nonumber\\
\partial_t E+\frac{1}{r^2}\partial_r[r^2u(E+P)]
 &=-\rho u g-\frac{1}{r^2}\partial_r(r^2q).
\label{eq:euler}
\end{align}
The three lines express conservation of SIDM mass, radial momentum, and total fluid energy, respectively.  Here $r$ is radius; $g$ is the total inward gravitational acceleration generated by the SIDM, compact baryonic component, and black hole; and $q$ is the conductive heat flux carried by SIDM self-interactions.  At the innermost boundary, SIDM reaching the black hole is removed from the fluid and added to its mass $M_{\rm BH}$; the resulting SIDM accretion rate includes the resolved inward flow and local collisional contribution.  Both are calculated from the evolving SIDM solution rather than prescribed.  The detailed transport closure, central-sink prescription, and diagnostics are given in the Supplemental Material.

We model only the net central concentration of the gas: $M_b=fM_h$ is the inflowing mass that forms a gaseous nucleus, not the halo's full cosmic baryon allotment $f_{\rm cos}M_h$ \cite{Planck2020}.  High-redshift merger simulations motivate values from $f\sim0.003$ to $0.03$ \cite{Mayer2014DirectCollapse,Mayer2023CosmoDCBH,Zwick2022MergerDCBH}.  Rather than a shell, the gas is a spherical Hernquist profile \cite{Hernquist1990} whose fixed mass begins with half-mass radius $0.1R_{\rm vir}$ and contracts smoothly to $R_{b,f}$ over $t_c=1$ Myr.  This clock represents final nuclear assembly: idealized calculations form a parsec-scale nucleus within a few $10^4$ yr of core coalescence, and a cosmological calculation shows a $4$ pc nucleus about $1$ Myr after the final merger, although the preceding merger lasts tens of Myr \cite{Mayer2014DirectCollapse,Mayer2023CosmoDCBH}.  We adopt $R_{b,f}=1$ pc and test 3 and 10 pc; these are phenomenological scales motivated by the simulations, not fitted gas-dynamical outputs.  The imposed potential captures the gravitational effect of baryonic contraction \cite{Blumenthal1986AdiabaticContraction,Gnedin2004AdiabaticContraction} without simulating the gas flow itself.

We solve Eqs.~(\ref{eq:euler}) for NFW halos at redshift $z=8$ with concentration $c=3$ \cite{NavarroFrenkWhite1997,BarkanaLoeb2001,DuttonMaccio2014}.  The parameter grid uses $M_h=10^8$, $10^9$, $10^{10}$, and $10^{11}M_\odot$; $f=0.003$, $0.01$, and $0.03$; $M_{\rm seed}=10^2$, $10^3$, and $10^4M_\odot$; and $\sigma/m=0.3$, 1, 3, and $10\,{\rm cm^2\,g^{-1}}$ \cite{Vogelsberger2012SIDM,Rocha2013SIDM,TulinYu2018SIDM}.  The full parameter atlas and numerical comparisons are given in the Supplemental Material.

\begin{figure*}[t]
\centering
\includegraphics[width=0.99\textwidth]{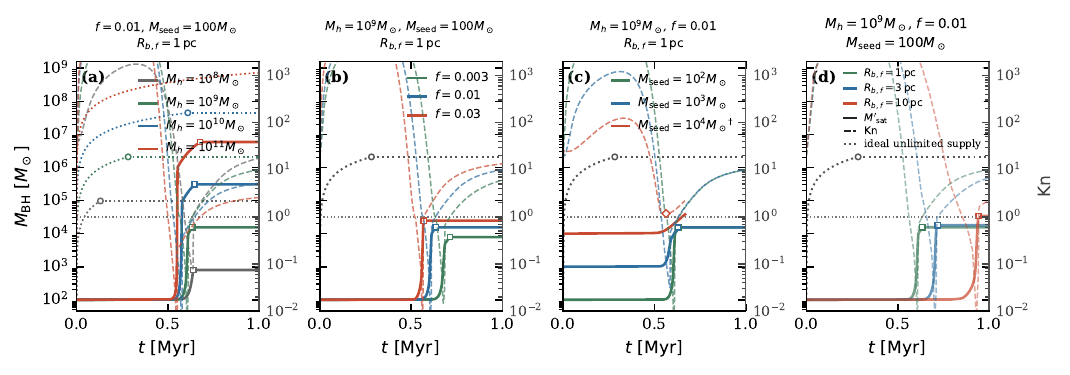}
\caption{\label{fig:growth}Black-hole growth during the prescribed $1$ Myr assembly of the gaseous nucleus.  Solid curves follow numerical $M_{\rm BH}(t)$ to the final upward ${\rm Kn}=1$ crossing, $t_{\rm off}$; squares mark $M'_{\rm sat}\equiv M_{\rm BH}(t_{\rm off})$.  Their horizontal continuation is only a plotting convention: slower growth after $t_{\rm off}$ is shown in the Supplemental Material.  Dashed curves show ${\rm Kn}$.  The dagger and diamond mark the near-critical $10^4M_\odot$ case, whose minimum ${\rm Kn}=1.17$ does not cross unity.  Dotted tracks and open circles show the ideal unlimited-supply SIDM model.}
\end{figure*}

\noindent\textbf{\emph{Effective collisional endpoints.---}}
At early times the seed mass changes little and ${\rm Kn}\gg1$, so the dark matter is too weakly collisional to sustain efficient fluid accretion.  During the assembly and contraction of the gaseous nucleus, its deepening potential drives SIDM inward, building a dense local reservoir whose evolution is coupled to collisional heat transport.  Figure~\ref{fig:growth} therefore focuses on $0\leq t\leq t_c=1$ Myr, from the onset of the prescribed inflow to completion of the gaseous nucleus.  In most cases, ${\rm Kn}$ falls below unity and the seed enters a brief phase of rapid growth while the nucleus is still contracting.  When ${\rm Kn}$ rises back to unity, we define the effective saturation mass
\begin{equation}
M'_{\rm sat}\equiv M_{\rm BH}(t_{\rm off}),
\qquad
{\rm Kn}(t_{\rm off})=1,
\end{equation}
where $t_{\rm off}$ is the final upward crossing.  This marks the end of the efficient collisional episode, not a terminal black-hole mass; subsequent, less collisionally coupled inflow can still add substantial mass.

Figure~\ref{fig:growth}(a) shows that increasing $M_h$ strongly raises $M'_{\rm sat}$, because more SIDM is accumulated before the collisional endpoint.  Panel (b) shows a similar but milder trend with $f$: a larger inflowing gas mass strengthens the central compression and triggers the endpoint earlier.  Panel (c) shows weak dependence on seed mass for $10^2$--$10^3M_\odot$, while the $10^4M_\odot$ seed is near-critical and does not trigger the burst.  Panel (d) shows a smaller, non-monotonic dependence on the final radius: in these runs larger $R_{b,f}$ gives a later and somewhat larger endpoint, so the time-dependent supply matters more than a simple compactness ranking.

The evolving reservoir therefore feeds a substantial burst, but the collisional episode ends well before the seed reaches the formal ideal mass, typically by about two orders of magnitude.  For the fiducial case, the interval from the first ${\rm Kn}<1$ sample to the upward crossing is about $0.07$ Myr, shorter than the ideal $\tau_{\rm sat}\simeq0.28$ Myr measured from $t=0$.  This difference in clock origins does not imply super-Bondi accretion: the burst starts only after baryonic contraction has assembled the local reservoir, and its mean rate remains below the ideal unlimited-supply rate.  The endpoint therefore marks loss of collisionality, not exhaustion of the whole SIDM supply.  Horizontal segments in Fig.~\ref{fig:growth} are only a plotting convention; untruncated histories and the supply-budget analysis are given in the Supplemental Material.

\noindent\textbf{\emph{Discussion and conclusion.---}}
The calculations identify a simple physical sequence.  Gas inflow first deepens the nuclear potential; SIDM heat transport and gravity then concentrate a finite dark reservoir; finally, a pre-existing seed consumes that reservoir at the rate supplied by the evolving solution.  Rapid growth is therefore an outcome of baryonic nuclear assembly, not an accretion law imposed on the black hole.  The same sequence also explains why the outcome need not be direct collapse.  A seed provides an absorbing center while the SIDM reservoir is still nonrelativistic; whether a gas core, an SIDM core, or the pre-existing seed reaches a black-hole-forming state first depends on their relative collapse and accretion clocks \cite{Mayer2014DirectCollapse,Mayer2023CosmoDCBH,FengYuZhong2021Seed,FengYuZhong2022RelativisticSIDM}.

This is deliberately a mechanism test.  Spherical symmetry replaces a rotating, fragmenting nucleus by its monopole, and the prescribed potential omits feedback and star formation \cite{Mayer2014DirectCollapse,Mayer2023CosmoDCBH}.  A constant $\sigma/m$ is likewise a benchmark for potentially velocity-dependent SIDM microphysics \cite{TulinYu2018SIDM}.  Nonetheless, the calculation establishes the essential point: gas-inflow-driven local SIDM contraction can feed an existing seed efficiently enough to make a massive black hole, while the finite reservoir produces growth histories qualitatively different from the ideal unlimited supply.  This channel complements direct-collapse and gas-accretion scenarios by turning the coupling between baryonic nuclear compaction and SIDM transport into a concrete black-hole growth pathway.

\begin{acknowledgments}
YR acknowledges support from the CAS Pioneer Hundred Talents Program (Category B), the NSFC grants 12522302, 12673017 and 12273037, and the USTC Research Funds of the Double First-Class Initiative.
\end{acknowledgments}

\bibliography{refs_v23}

\end{document}